\documentclass[superscriptaddress,showpacs,preprintnumbers,amsmath,amssymb]{revtex4}
\usepackage{graphicx}
\usepackage{dcolumn}
\usepackage{bm}
\usepackage{latexsym}
\begin{document}
\title{Distribution of dwell times of a ribosome:\\ effects of infidelity, kinetic proofreading and ribosome crowding}
\author{Ajeet K. Sharma}
\author{Debashish Chowdhury{\footnote{Corresponding author(E-mail: debch@iitk.ac.in)}}}
\affiliation{Department of Physics, Indian Institute of Technology,
Kanpur 208016, India.}
\begin{abstract} 
Ribosome is a molecular machine that polymerizes a protein where 
the sequence of the amino acid residues, the monomers of the 
protein, is dictated by the sequence of codons (triplets of 
nucleotides) on a messenger RNA (mRNA) that serves as the template. 
The ribosome is a molecular motor that utilizes the template mRNA 
strand also as the track. Thus, in each step the ribosome moves 
forward by one codon and, simultaneously, elongates the protein by 
one amino acid. We present a theoretical model that captures most 
of the main steps in the mechano-chemical cycle of a ribosome. 
The stochastic movement of the ribosome consists of an alternating 
sequence of pause and translocation; the sum of the durations of a 
pause and the following translocation is the time of dwell of the 
ribosome at the corresponding codon. We derive the analytical 
expression for the distribution of the dwell times of a ribosome in 
our model. Whereever experimental data are available, our theoretical 
predictions are consistent with those results. We suggest appropriate 
experiments to test the new predictions of our model, particularly, 
the effects of the quality control mechanism of the ribosome and 
that of their crowding on the mRNA track. 
\end{abstract}
\maketitle
\section{Introduction} 
\label{sec-introduction}

The primary structure of a protein consists of a sequence of amino acid 
residues linked together by peptide bonds. Therefore, a protein is also 
referred to as a polypeptide which is essentially a linear hetero-polymer, 
the amino acid residues being the corresponding monomers.
The sequence of the amino acid residues in a polypeptide is dictated by 
that of the codons, each of which is a triplet of nucleotides, on 
a messenger RNA (mRNA) that serves as the template for the polypeptide. 
The actual polymerization of each protein from the corresponding mRNA 
template is carried out by a macromolecular machine called ribosome 
\cite{spirinbook,spirin02,spirin09} 
and the process is referred to as {\it translation} (of genetic code). 
The polymerization of protein takes places in three stages that are 
identified as {\it initiation}, {\it elongation} (of the protein) and 
{\it termination}. 

The ribosome also utilizes the template mRNA as a track for its own 
movement; it steps forward by one codon at a time while, simultaneously, 
it elongates the growing polypeptide by one amino acid monomer. Therefore, 
ribosome is also regarded as a motor that, like other molecular motors, 
takes input (free-) energy from the hydrolysis of a nucleotide 
tri-phosphate to move along a filamenous track \cite{cross97}. In fact, 
a ribosome hydrolyzes two molecules of Guanosine tri-phosphate (GTP) 
to move forward by one codon.

Enormous progress has been made in the last decade in the fundamental 
understanding of the structure, energetics and kinetics of ribosomes 
\cite{ramanobel,adanobel,steitznobel,steitznature,ramanature,ramacell,frank06a,frank06b,frank09,frank10}.
Recent single molecule studies of ribosomes 
\cite{marshall08,munro08,munro09,blanchard09,blanchard04,uemura10,aitken10} 
has thrown light on its operational mechanism. 

In single molecule experiments, it has been observed that a ribosome 
steps forward in a stochastic manner; its stepping is characterized 
by an alternating sequence of pause and translocation. The sum of the 
durations of a pause and the following translocation defines the time 
of a dwell of the ribosome at the corresponding codon. The time of 
dwell of a ribosome varies randomly from one codon to another. This 
randomness arises from two different sources: 
(i) {\it intrinsic} fluctuations associated with the Brownian forces 
as well as the low of concentrations of the molecular species involved 
in the chemical reactions, and (ii) {\it extrinsic} fluctuations 
caused by the inhomogeneities of the sequence of nucleotides on the 
template mRNA \cite{buchan07}. Because of the sequence inhomogeneity 
of the mRNA templates used by Wen et al. \cite{wen08}, the dwell time 
distribution (DTD) measured in their single-molecule experiment reflects 
a combined effect of the intrinsic and extrinsic fluctuations on the 
dwell time. In contrast, in this paper we focus on situations (to be 
explained in detail in the following sections) where the randomness 
of the dwell times arises exclusively from {\it intrinsic} fluctuations.

The probability density $f(t)$ of the dwell times of a ribosome, 
measured in single-molecule experiments \cite{wen08}, has been 
compared with the corresponding data obatined from computer 
simulations \cite{wen09}. It has been claimed that the data do 
not fit a single exponential thereby indicating the existence of 
more than one rate-limiting step in the mechano-chemical cycle of 
each ribosome. In fact, the best fit to the simulation data was 
achieved with five different rate-determining steps \cite{wen09}.  

A systematic analytical derivation of the DTD of the ribosomes was 
presented recently \cite{gccr} from a kinetic theory of translation 
\cite{basu07} that also involves essentially five steps in the 
mechano-chemical cycle of a ribosome. However, the model developed 
in ref.\cite{basu07} ignores some of the key features of the 
mechano-chemical cycle of an individual ribosome during the elongation 
stage. For example, a ribosome deploys an elaborate proofreading 
mechanism to select the correct amino acid dictated by the template 
(and to reject the incorrect ones) to ensure high translational 
fidelity. Nevertheless, occasionally, translational errors take place. 
In this paper we extend Basu and Chowdhury's original model 
\cite{basu07} by capturing the processes of proofreading and allowing 
for the possibility of imperfect fidelity of translation. Moreover, 
the identification of the mechano-chemical states as well as the 
nature of the transitions among these states is revised in the light 
of the empirical facts established in the last couple of years. 
Using this revised and extended kinetic model of translation 
\cite{sharmathesis}, we {\it analytically} calculate the probability 
density $f(t)$ of the dwell times of ribosomes. 

The DTD derived in this paper is qualitatively similar to that observed 
by Wen et al. in their single molecule experiments \cite{wen08}. 
However, because of the sequence inhomogeneity of the template mRNA 
used in the experiment of Wen et al.\cite{wen08}, their data cannot 
be compared quantitatively with our analytical expression for $f(t)$. 
Therefore, we propose a concrete experimental set up required for a 
quantitative testing of our theoretical predictions.  However, till 
such experiments are actually carried out, the results reported here 
will continue to provide insight into the mechanistic  origin of the 
qualitative features of the DTD. Moreover, attempts are being made to 
extend our model to capture sequence inhomogeneities of mRNAs and to 
calculate the corresponding DTDs of ribosomes {\it analytically}.

Interestingly, the inverse of the mean dwell time is the average 
velocity $V$ of the ribosome motor. Since, in our model, the ribosome 
is not allowed to back track on the mRNA template, $V$ is also the 
mean rate of elongation of the polypeptide. We show that $V$ satisfies 
an equation that resembles the Michaelis-Menten equation for the average 
rate of simple enzymatic reactions \cite{dixon}.

It is well known  that most often a large number of ribosomes
simultaneously move on the same mRNA track each polymerizing
one copy of the same protein. This phenomenon is usually referred
to as ribosome traffic because of its superficial similarity with
vehicular traffic on highways
\cite{macdonald68,macdonald69,lakatos03,shaw03,shaw04a,shaw04b,chou03,chou04,dong1,dong2,cook,ciandrini,basu07,mitarai08}
In this paper we also report the effect of steric interactions of 
the ribosomes in ribosome traffic on their DTD.

\section{The model} 
\label{sec-model}

We develope here a theoretical model for the polymerization of a 
protein by ribosomes using an artificially synthesized mRNA 
template that consists of a {\it homogeneous} sequence (i.e., 
all the codons of which are identical). In the surrounding medium, 
{\it two} species of amino-acids are available only one of which is 
the correct one according to the genetic code. The ribosome deploys 
a quality control system that rejects the amino acid monomer if it 
is incorrect. If this quality control system never fails, perfect 
fidelity of translation would result in a homopolymer whose 
constituent amino acid monomers are all identical. However, 
occasionally, wrong amino acid monomer escapes the quality control 
system. Such translational errors, whereby a wrong amino acid 
monomer is incorporated in the elongating protein, results in a 
{\it hetero-polymer}. We'll study the effects of the quality control 
system on the DTD of the ribosomes.

\begin{figure}[t]
\begin{center}
\includegraphics[angle=-90,width=0.6\columnwidth]{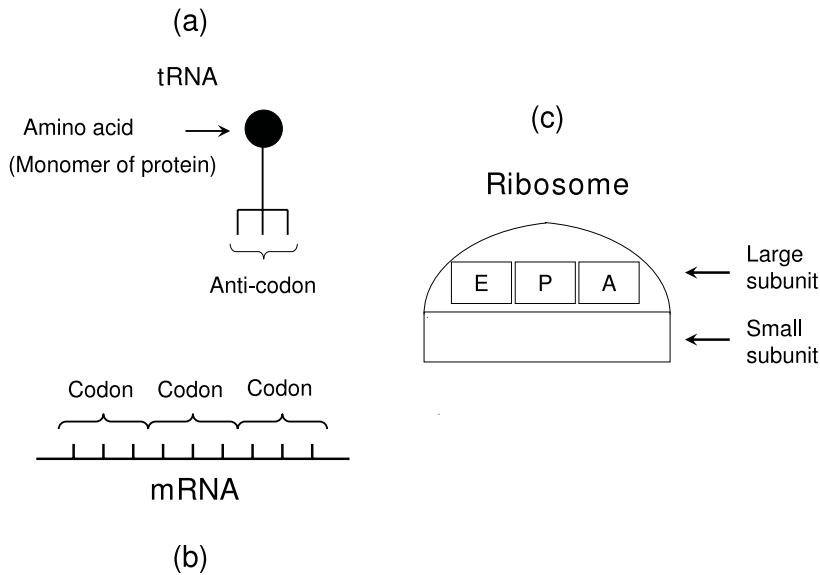}
\end{center}
\caption{Cartoons representing the main molecules and machinery 
involved in the process of {\it translation}. 
}
\label{fig-parts}
\end{figure}

A ribosome consists of two interconnected subunits which are 
designated as ``large'' and ``small'' (see fig.\ref{fig-parts}). 
The small subunit binds with the mRNA track and decodes the 
genetic message of the codon whereas the polymerization of the 
protein takes place in the large subunit. The operations of the 
two subunits are coordinated by a class of adapter molecules, 
called tRNA (see fig.\ref{fig-parts}). One end of a tRNA helps 
in the decoding process by matching its anti-codon with the codon 
on the mRNA while its other end carries an amino acid; 
in this form the complex is called an amino-acyl tRNA (aa-tRNA).

\begin{figure}[t]
\begin{center}
\includegraphics[angle=-90,width=0.6\columnwidth]{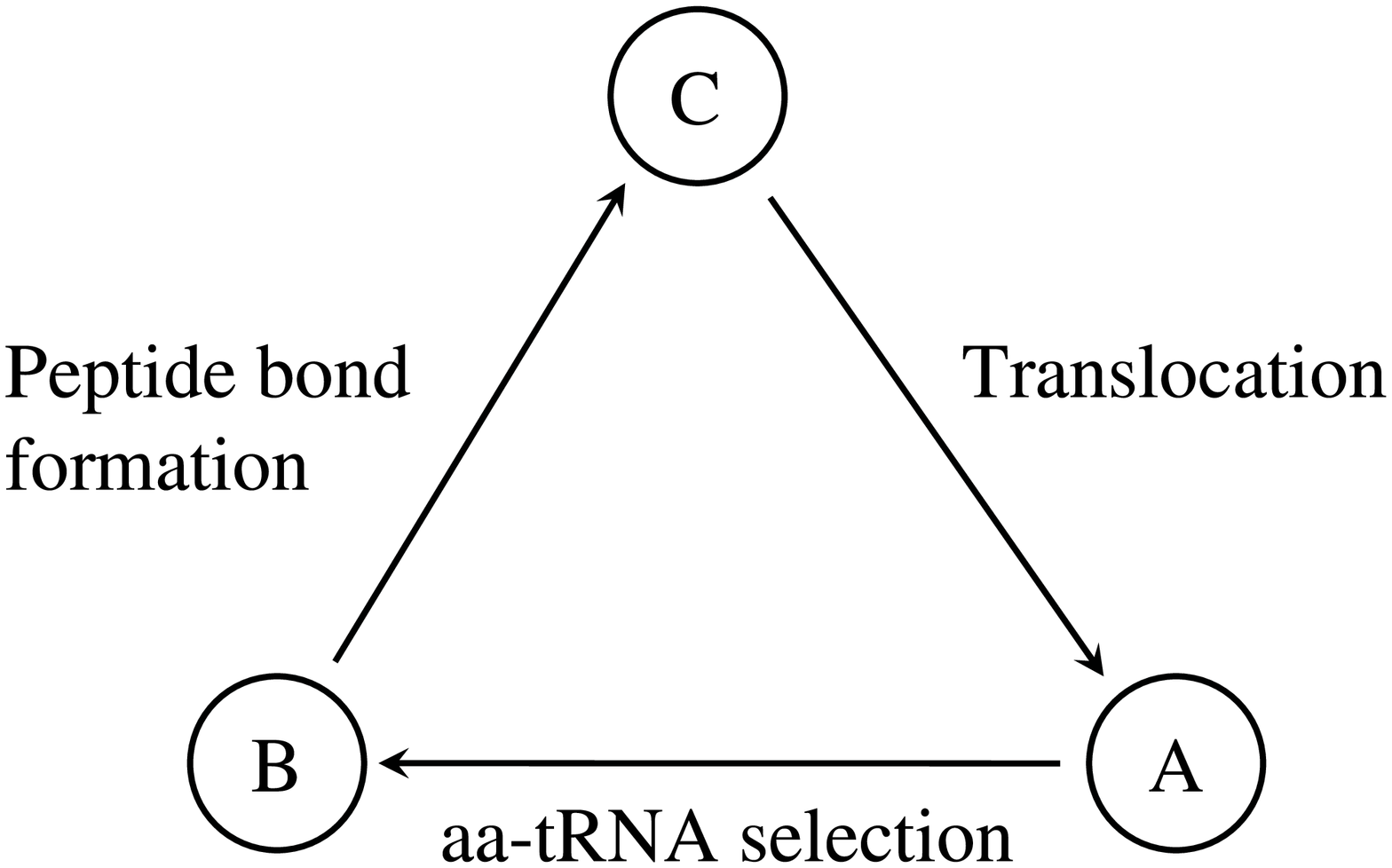}
\end{center}
\caption{Pictoral depiction of the three main stages in the 
chemo-mechanical cycle of a single ribosome (see the text for 
details).
}
\label{fig-3state}
\end{figure}

\begin{figure}[t]
\begin{center}
\includegraphics[angle=-90,width=0.6\columnwidth]{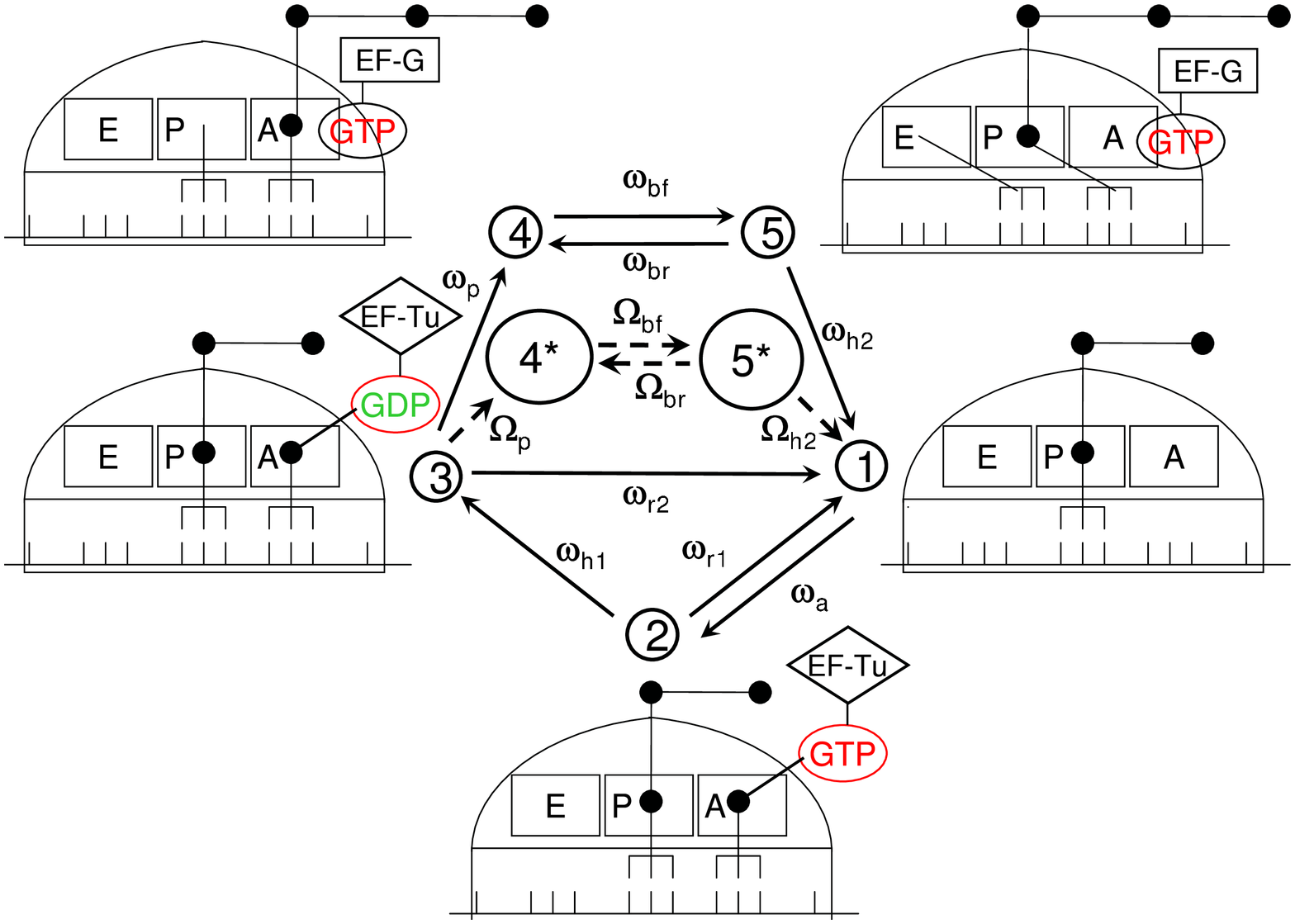}
\end{center}
\caption{Pictoral depiction of the main steps in the chemo-mechanical 
cycle of a single ribosome (see the text for details).
}
\label{fig-5state}
\end{figure}

The three main stages in each mechano-chemical cycle of a ribosome, 
shown in fig.\ref{fig-3state}, are as follows: (i) selection of the 
cognate (i.e., correct) aa-tRNA, (ii) formation of the peptide bond 
between the amino acid brought in by the selected aa-tRNA and the 
elongating protein, (iii) translocation of the ribosome by one codon 
on its track. However, some of these steps consist of important 
sub-steps. Moreover, the aa-tRNA selected (erroneously) by the 
ribosome may not be the cognate tRNA; this leads to a branching of 
pathways. Such branching, in turn, gives rise to the possibility of 
more than one cyclic pathway for a ribosome in a given cycle. 

Let us begin with the state labelled by ``1''; both the E and A sites 
are empty while the site P is occupied as shown in fig.\ref{fig-5state}. 
An incoming aa-tRNA molecule, bound to an elongation factor EF-Tu, 
occupies the site A; the resulting state is labelled by ``2''. The 
transition $1 \to 2$ takes place at the rate $\omega_{a}$. However, 
not all incoming aa-tRNA molecules are automatically selected by the 
ribosome. In order to ensure translational {\it fidelity} 
\cite{rodnina01,rodnina06,ogle05,zaher09}
the ribosome deploys a quality control mechanism to ensure that the 
aa-tRNA selected is, indeed, cognate (i.e., carries the correct amino 
acid as dictated by the mRNA template). A non-cognate tRNA is rejected 
on the basis of codon-anticodon mismatch; the corresponding transition 
$2 \to 1$ takes place at the rate $\omega_{r1}$. However, if the aa-tRNA 
is not a non-cognate one, the GTP molecule bound to it is then hydrolyzed 
to GDP causing the irreversible transition $2 \to 3$ at the rate 
$\omega_{h1}$. At this stage, kinetic proofreading can reject the 
tRNA if it is a near-cognate one; the transition $3 \to 1$ captures 
this phenomenon and takes place at the rate $\omega_{r2}$. In our model 
we do not explicitly treat those sub-steps in which EF-TU and the 
products of the hydrolysis of GTP leave the ribosome. We'll later 
utilize the fact that $\omega_{a}$ is proportional to the concentration 
of the aa-tRNA molecules in the surrounding medium.

If the selected aa-tRNA is not rejected, it does not necessarily imply 
that it is cognate because, occasionally, even non-cognate or near-cognate 
aa-tRNA escapes the ribosomes quality control mechanism. The amino acid 
supplied by the selected aa-tRNA is then linked to the growing 
polypeptide by a peptide bond; this peptidyl transferase activity of 
the ribosome elongates the polypeptide by one monomer. The correct amino 
acid is incorporated at a rate $\omega_p$ whereas a wrong amino acid is 
incorporated at a rate $\Omega_p$ thereby giving rise to the two different 
branches of the mechano-chemical cycle of the ribosome; the third branch 
$3 \to 1$,  however, ends a ``futile'' cycle. The only difference between 
the states $4$ and $4^{*}$ is that the last amino acid in the polypeptide 
is correct in $4$ but incorrect in $4^{*}$. While the polypeptide gets 
elongated by one amino acid, a fresh molecule of GTP enters bound with 
an elongation factor EF-G. The arrival of GTP-bound EF-G is not treated 
explicitly in our model. Alternatively, $\omega_{p}$ (and $\Omega_{p}$) 
is an effective rate constant that accounts for both the polypeptide 
elongation and the arrival of the GTP-bound EF-G.

Next, spontaneous Brownian (relative) rotation of the two subunits of 
the ribosome coincides with the back-and-forth transition between the 
so-called ``classical'' and ``hybrid'' configurations of the two tRNA 
molecules. 
\cite{noller89,noller98,noller02,noller07,noller08,frank00,frank03,frank07,pan07,moran08, shoji09} 
In the classical configuration, both ends of the two tRNA molecules 
correspond to the locations of $P$ and $A$ sites. In contrast, in the 
hybrid configuration the ends of tRNA molecules interacting with the 
large subunit are found at the locations of $E$ and $P$ sites, 
respectively. The forward transition $4 \to 5$ takes place at the 
rate $\omega_{bf}$ along the correct branch (and at the rate 
$\Omega_{bf}$ along the wrong branch) whereas the reverse transition 
$5 \to 4$ takes place at the rate $\omega_{br}$ along the correct 
branch (and $\Omega_{br}$ along the wrong branch). The reversible 
transition $4 \rightleftharpoons 5$, which is caused by spontaneous 
Brownian fluctuations, does not need any free energy input from GTP 
hydrolysis.  

Finally the hydrolysis of GTP drives the irreversible transition 
$5 \to 1$ which involves the translocation of the ribosome on its 
track by one codon and, simultaneously, that of the tRNAs inside 
the ribosome by one binding site following which the deacetylated 
(i.e., denuded of amino acid) tRNA exits from the E site. The rate 
of this transition is $\omega_{h2}$ or $\Omega_{h2}$ depending on 
whether the system is following the correct branch or the wrong 
branch. In our model, the transition $5 \to 1$ captures the hydrolysis 
of GTP, departure of the products of hydrolysis alongwith EF-G as well 
as the final exist of the deacetylated tRNA molecule from the E site.

\begin{figure}[t]
\begin{center}
\includegraphics[angle=-90,width=0.6\columnwidth]{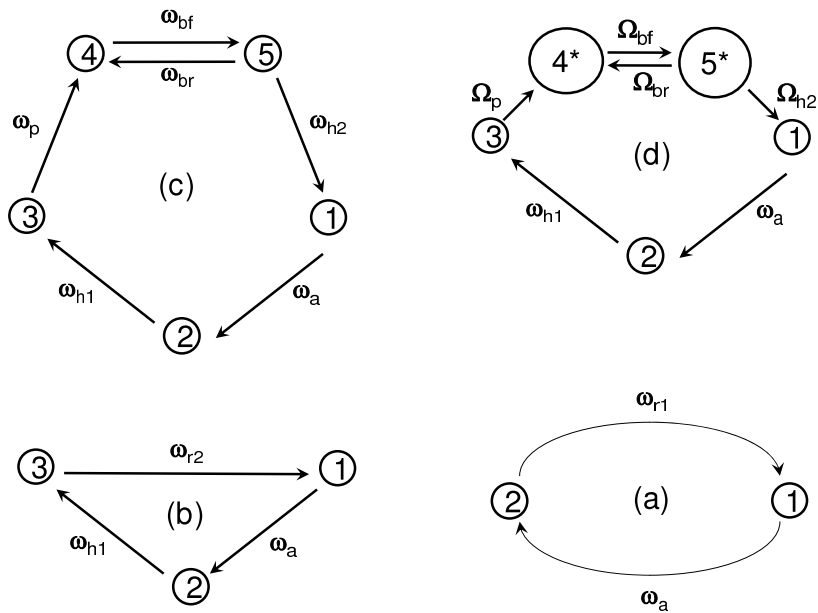}
\end{center}
\caption{The full mechano-chemical kinetic of a ribosome during the 
elongation state, as modelled in Fig.\ref{fig-5state}, can be 
regarded as a composite of at least four different cycles shown 
in (a)-(d). Among these, both (a) and (b) are unproductive in the 
sense that these lead to neither elongation of the polypeptide nor 
forward stepping of the ribosome. However, (b) is ``futile'' because 
it dissipates the free energy input from the hydrolysis of a GTP 
molecule whereas no fuel is wasted in (a) . In contrast, both (c) 
and (d) lead to elongation of the polypeptide and forward stepping 
of the ribosome by hydrolysing two molecules of GTP; however, (c) 
incorporates the correct amino acid whereas (d) incorporated a wrong 
amino acid into the growing polypeptide. 
}
\label{fig-ribo4cycles}
\end{figure}

The rate constant $\omega_{a}$ can be made identical for both species 
of amino acid monomers by maintaining their concentrations in the 
medium appropriately. The rate constant $\omega_{r1}$ refers only to 
the wrong amino acids (i.e., non-cognate tRNA) because, we assume, 
cognate tRNA is not rejected at all. Since the transition $2 \to 3$ 
accounts for  the hydrolysis of a GTP molecule by the GTPase EF-Tu, 
the corresponding rate constant is $\omega_{h1}$, irrespective of 
the identity of the attached aa-tRNA. In kinetic proofreading, the 
rate of rejection of the non-cognate tRNA is much higher than that 
of a cognate tRNA. For the sake of simplicity, we assume that the 
cognate tRNA is not rejected at all. Therefore, the rate constant 
$\omega_{r2}$ referers to the rejection of only non-cognate tRNA. 
For the remaining steps of the mechano-chemical cycle, the rate 
constants are $\omega_{p}$, $\omega_{bf}$, $\omega_{br}$ and 
$\omega_{h2}$ provided a cognate tRNA has been selected finally. 
On the other hand, the corresponding rate constants are 
$\Omega_{p}$, $\Omega_{bf}$, $\Omega_{br}$ and $\Omega_{h2}$, 
respectively, if a non-cognate tRNA escapes rejection by the quality 
control system of the ribosome. Since the last step involves not only 
hydrolysis of GTP by the GTPase EF-G, but also translocation of the 
ribosome and the tRNA molecules, the rate constants $\omega_{h2}$ and 
$\Omega_{h2}$ need not be equal, in general.

Thus our model is an extension of the generic models for molecular 
motors based on stochastic chemical kinetic approach which was 
pioneered by Fisher and Kolomeisky \cite{fishkolo,kolorev}.
Following Hill \cite{hillbook} the mechano-chemical kinetics of a 
ribosome in our model can be regarded as a composite of the four 
cycles shown in Fig.\ref{fig-ribo4cycles}. 

All the numerical data generated from the analytical expressions 
for the graphical plots have the following set of values of the 
rate constants (except in the figures where $f(t)$ has been 
plotted for several different values of the parameters 
$\omega_{p}$ and $\omega_{r2}$) : 
$\omega_{a} = 25$s$^{-1}$, 
$\omega_{r1} = 10$s$^{-1}$,
$\omega_{h1} = 25$s$^{-1}$,
$\Omega_{p} = 40$s$^{-1}$, 
$\omega_{bf} = \omega_{br} = 25$s$^{-1}$, 
$\Omega_{bf} = \Omega_{br} = 10$s$^{-1}$, 
$\omega_{h2} = 25$s$^{-1}$,
$\Omega_{h2} = 10$s$^{-1}$.
The value of $\omega_{a}= 25$s$^{-1}$ is identical to that used in our 
earlier papers (see \cite{basu07,gccr} and the references therein). 
For the purpose of plotting our results, the magnitude of the other 
parameters have been chosen to be comparable to that of $\omega_{a}$. 
The values of some of the parameters, and the ranges over which some 
of these have been varied, may be unrealistically high or low. But, 
this  deliberate choice has been motivated by our intention to 
demonstrate graphically the interplay of various kinetic processes in 
translation.

\section{Dwell time distribution} 
\label{sec-dtd}

For the convenience of mathematical calculations, we assume that, after 
reaching the state $5$ (or $5^*$) at location $j$ the system makes a 
transition to a hypothetical state $\widetilde{1}$ at location $j+1$ which, 
then, relaxes to the state $1$, at the same location, at the rate 
$\delta$. At the end of the calculation our model is recovered by setting 
$\delta \to \infty$. 

Suppose $P_{\mu}(j,t)$ denote the probability at time $t$ that the 
ribosome is in the ``chemical'' state $\mu$ and is decoding the 
$j$-th codon. Let us use the symbol $\widetilde{P}_1(j+1,t)$ to 
denote the probability of finding the ribosome in the hypothetical 
state $\widetilde{1}$ at time $t$ while decoding the $j+1$-th codon. 
The time taken by the ribosome to reach the state $\widetilde{1}$ at 
$j+1$, starting from the initial state $1$ at $j$, defines its time 
of dwell at the $j$-th codon. Since, in this context, all the 
``chemical'' states, except $\widetilde{1}$ refer to the $j$-th codon 
while $\widetilde{1}$ corresponds to the $j+1$-th codon, from now onwards 
we drop the site index $j$ (and $j+1$) to keep the notations simple. 

The master equations governing the time evolution of the probabilities 
$P_{\mu}(t)$ can be written as
\begin{equation}
\frac{dP_1(t)}{dt} = -\omega_a P_1(t)+\omega_{r1} P_2(t)+ \omega_{r2} P_3(t)
\label{eq-m1}
\end{equation}
\begin{equation}
 \frac{dP_2(t)}{dt}=\omega_a P_1(t)-(\omega_{r1}+\omega_{h1}) P_2(t)
\label{eq-m2}
\end{equation}
\begin{equation}
 \frac{dP_3(t)}{dt}=\omega_{h1} P_2(t)- (\omega_p+\Omega_p+\omega_{r2})P_3(t)
\label{eq-m3}
\end{equation}
\begin{equation}
 \frac{dP_4(t)}{dt}=\omega_p P_3(t)-\omega_{bf} P_4(t)+\omega_{br} P_5 (t)
\label{eq-m4}
\end{equation}
\begin{equation}
 \frac{dP_5(t)}{dt}=\omega_{bf} P_4(t)-(\omega_{h2}+\omega_{br}) P_5(t)
\label{eq-m5}
\end{equation}
\begin{equation}
 \frac{dP_4^*(t)}{dt}=\Omega_p P_3(t)-\Omega_{bf} P_4^*(t)+\Omega_{br} P_5^* (t)
\label{eq-m6}
\end{equation}
\begin{equation}
 \frac{dP_5^*(t)}{dt}=\Omega_{bf} P_4^*(t)-(\Omega_{h2}+\Omega_{br}) P_5^*(t)
\label{eq-m7}
\end{equation}
\begin{equation}
\frac{d\widetilde{P_1}(t)}{dt}=\omega_{h2}P_5(t)+\Omega_{h2}P_5^*(t)
\label{eq-m8}
\end{equation}
Because of the normalization condition
\begin{equation}
\sum_{\mu=1}^{5} P_{\mu}(t)+P_4^*(t)+P_5^*(t)+\widetilde{P_1}(t)=1
\end{equation}
not all the equations (\ref{eq-m1})-(\ref{eq-m8}) above are 
independent of each other. 

For the calculation of the dwell time, we impose the initial conditions  
\begin{equation}
P_1(0)=1, ~{\rm and}~P_2(0)=P_3(0)=P_4(0)=P_5(0)=P_4^*(0)=P_5^*(0)=\widetilde{P_1}(0)=0 
\label{eq-initial}
\end{equation}
Suppose, $f(t)$ is the probability density of the dwell times. Then, 
the probability of adding one amino acid to the growing polypeptide 
in the time interval between $t$ and $t+\Delta t$ is $f(t)\Delta t$ 
where 
\begin{equation}
f(t)=\dfrac{\Delta\widetilde{P_1}(t)}{\Delta t}=\omega_{h2}P_5(t)+\Omega_{h2}P_5^*(t)
\end{equation}
The calculation of the DTD  
\cite{chemla08,shaevitz05,liao07,linden07,fisher07,garai10} is 
essentially that of a distribution of first-passage times \cite{redner}. 

\begin{figure}[t]
\begin{center}
\includegraphics[angle=-90,width=0.6\columnwidth]{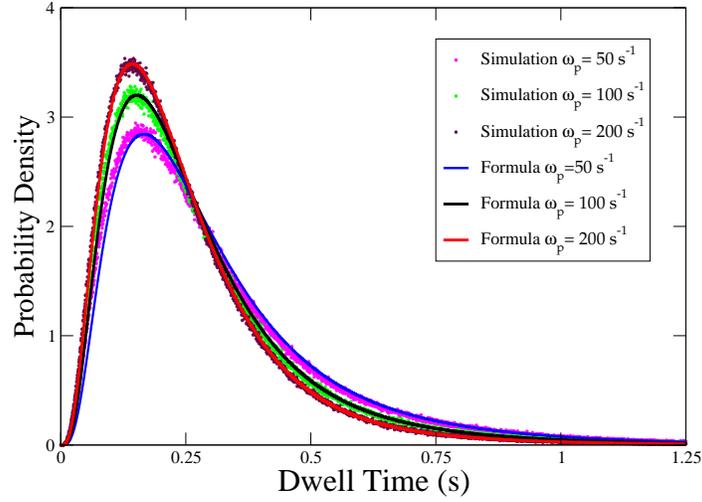}
\end{center}
\caption{The probability density of the dwell times are plotted for 
a few different values of the parameter $\omega_{p}$. 
}
\label{fig-ftomegap}
\end{figure}

\begin{figure}[t]
\begin{center}
\includegraphics[angle=-90,width=0.6\columnwidth]{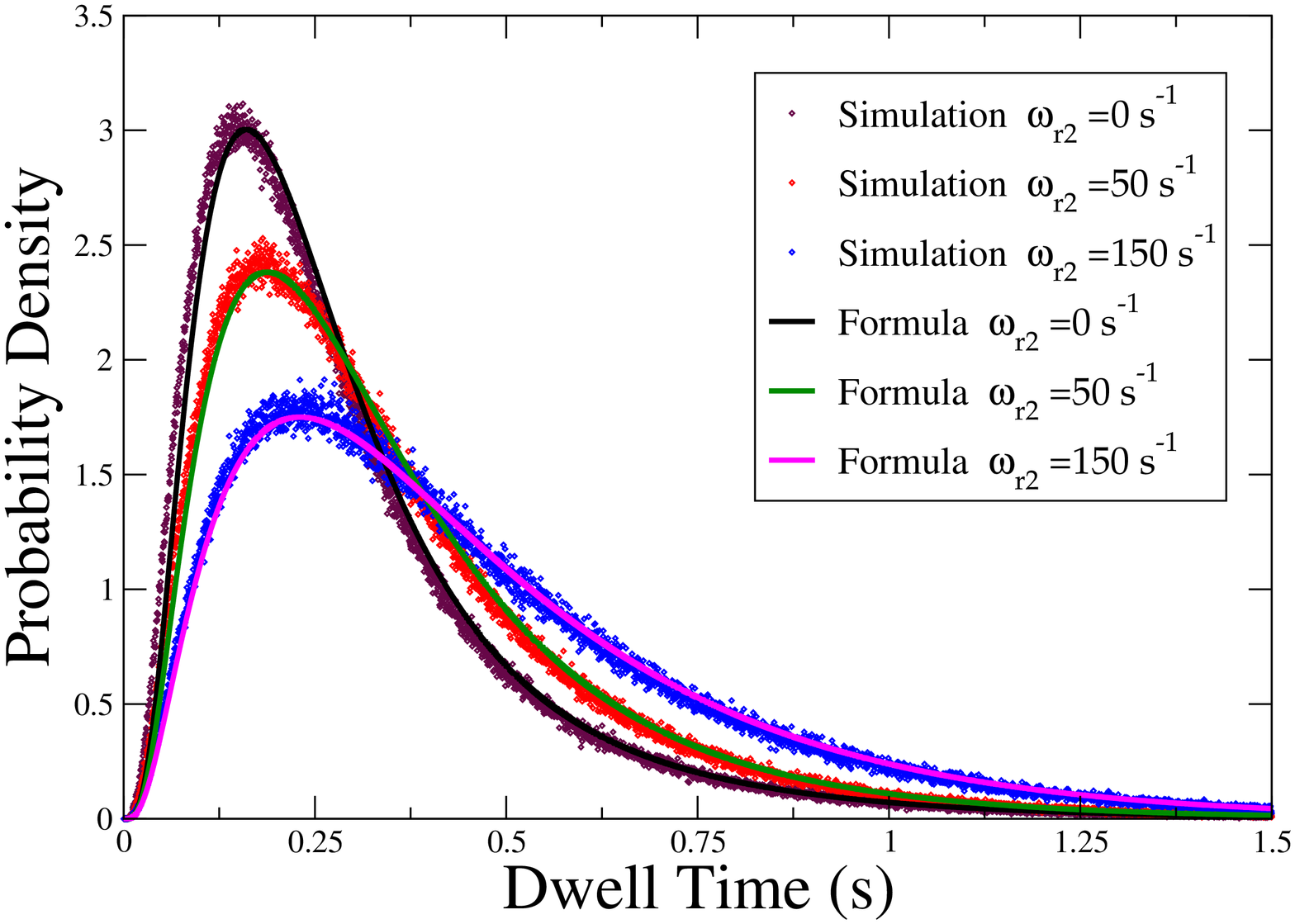}
\end{center}
\caption{The probability density of the dwell times are plotted for 
a few different values of the parameter $\omega_{r2}$. 
}
\label{fig-ftomegar2}
\end{figure}

\begin{figure}[t]
\begin{center}
\includegraphics[angle=-90,width=0.6\columnwidth]{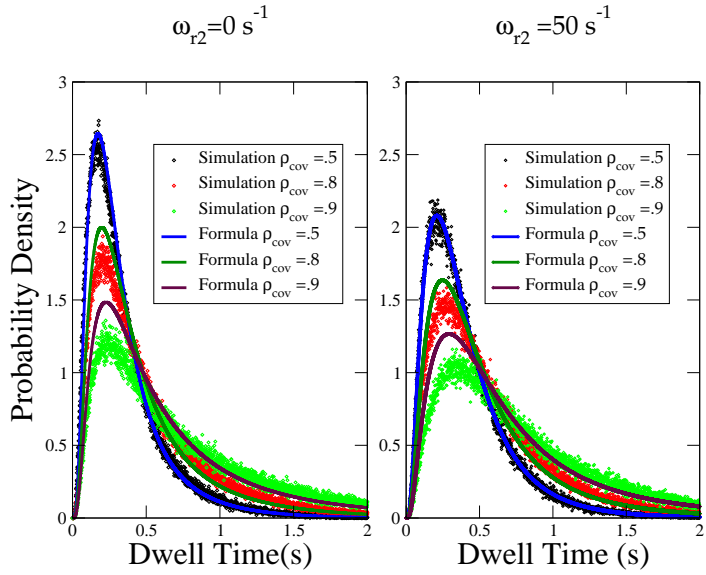}
\end{center}
\caption{The probability density of the dwell times are plotted for 
a few different values of the coverage density $\rho_{cov}$ of the 
ribosomes. 
}
\label{fig-ftrho}
\end{figure}

The {\it exact} probability density of the dwell times is given by 
\begin{eqnarray}
f(t)&=&\biggl[\dfrac{\omega_{h2}\omega_{bf}\omega_p\omega_{h1}\omega_a}{(\omega_1-\omega_2)(\omega_1-\omega_3)(\omega_1-\omega_4)(\omega_1-\omega_5)}\biggr]e^{-\omega_1 t} 
+\biggl[\dfrac{\Omega_{h2}\Omega_{bf}\Omega_p\omega_{h1}\omega_a}{(\omega_1-\omega_2)(\omega_1-\omega_3)(\omega_1-\Omega_4)(\omega_1-\Omega_5)}\biggr]e^{-\omega_1 t}  \nonumber\\
&+&\biggl[\dfrac{\omega_{h2}\omega_{bf}\omega_p\omega_{h1}\omega_a}{(\omega_2-\omega_3)(\omega_2-\omega_4)(\omega_2-\omega_5)(\omega_2-\omega_1)}\biggr]e^{-\omega_2 t} 
+\biggl[\dfrac{\Omega_{h2}\Omega_{bf}\Omega_p\omega_{h1}\omega_a}{(\omega_2-\omega_3)(\omega_2-\Omega_4)(\omega_2-\Omega_5)(\omega_2-\omega_1)}\biggr]e^{-\omega_2 t} \nonumber \\ 
&+&\biggl[\dfrac{\omega_{h2}\omega_{bf}\omega_p\omega_{h1}\omega_a}{(\omega_3-\omega_4)(\omega_3-\omega_5)(\omega_3-\omega_1)(\omega_3-\omega_2)}\biggr]e^{-\omega_3 t}
+\biggl[\dfrac{\Omega_{h2}\Omega_{bf}\Omega_p\omega_{h1}\omega_a}{(\omega_3-\Omega_4)(\omega_3-\Omega_5)(\omega_3-\omega_1)(\omega_3-\omega_2)}\biggr]e^{-\omega_3 t} \nonumber\\
&+&\biggl[\dfrac{\omega_{h2}\omega_{bf}\omega_p\omega_{h1}\omega_a}{(\omega_4-\omega_5)(\omega_4-\omega_1)(\omega_4-\omega_2)(\omega_4-\omega_3)}\biggr]e^{-\omega_4 t} 
+\biggl[\dfrac{\Omega_{h2}\Omega_{bf}\Omega_p\omega_{h1}\omega_a}{(\Omega_4-\Omega_5)(\Omega_4-\omega_1)(\Omega_4-\omega_2)(\Omega_4-\omega_3)}\biggr]e^{-\Omega_4 t} \nonumber \\
&+&\biggl[\dfrac{\omega_{h2}\omega_{bf}\omega_p\omega_{h1}\omega_a}{(\omega_5-\omega_1)(\omega_5-\omega_2)(\omega_5-\omega_3)(\omega_5-\omega_4)}\biggr]e^{-\omega_5 t} 
+\biggl[\dfrac{\Omega_{h2}\Omega_{bf}\Omega_p\omega_{h1}\omega_a}{(\Omega_5-\omega_1)(\Omega_5-\omega_2)(\Omega_5-\omega_3)(\Omega_5-\Omega_4)}\biggr]e^{-\Omega_5 t} \nonumber\\
\label{eq-ftfinal}
\end{eqnarray}
where $\omega_1$,$\omega_2$ and $\omega_3$ are solution of the cubic equation
\begin{eqnarray}
&&\omega^3 -\omega^2(\omega_{r1}+\omega_{h1}+\omega_a+\omega_{r2}+\omega_p+\Omega_p)+\omega(\omega_{h1}\omega_a+\omega_{r2}\omega_{r1}+\omega_{r2}\omega_{h1}+\omega_{r2}\omega_a\nonumber \\
&+&\omega_p\omega_{r1}+\omega_p\omega_{h1}+\omega_p\omega_a+\Omega_p\omega_{r1}+\Omega_p\omega_{h1}+\Omega_p\omega_a)-\Omega_p\omega_{h1}\omega_a+\omega_p\omega_{h1}\omega_a=0,
\label{eq-w123}
\end{eqnarray}
$\omega_4$ and $\omega_5$ are the solution of the quadratic equation
\begin{equation}
\omega^2-\omega(\omega_{h2}+\omega_{br}+\omega_{bf})+\omega_{h2}\omega_{bf}=0
\label{eq-w45}
\end{equation}
and $\Omega_4$ and $\Omega_5$ are the solution of the quadratic equation
\begin{equation}
\Omega^2-\Omega(\Omega_{h2}+\Omega_{br}+\Omega_{bf})+\Omega_{h2}\Omega_{bf}=0
\label{eq-w45star}
\end{equation}
Some of the details of this derivation are given in the appendix. 
Note that the problem of determining the rates $\omega_i$ 
($i=1,2,3,4,5$) in terms of the rate constants of the model is 
similar to that of expressing the normal modes of vibration 
of a set of coupled harmonic oscillators. It is the ``backward''  
reactions in the mechano-chemical cycle of our model which play 
the role of coupling of the harmonic oscillators. For example, 
if $\Omega_{p}= 0$ and simultaneously,
$\omega_{r1} = \omega_{r2} = \omega_{br} = 0$, the expressions 
for $\omega_{i}$ ($i=1,2,3,4,5$) reduce to the simple form 
$\omega_{1} = \omega_{a}$, $\omega_{2} = \omega_{h1}$, 
$\omega_{3} = \omega_{p}$, $\omega_{4} = \omega_{bf}$, and 
$\omega_{5} = \omega_{h2}$.

The distribution (\ref{eq-ftfinal}) is plotted in Fig.\ref{fig-ftomegap} 
for a few different values of the parameter $\omega_{p}$.  Note 
that 
\begin{equation}
\phi = \omega_{p}/(\omega_{p}+\Omega_{p}) 
\end{equation} 
is a measure of the fidelity of translation. Therefore, increasing 
$\omega_{p}$, keeping $\Omega_{p}$ fixed, enhances translational 
fidelity. Moreover, a higher $\omega_{p}$ also corresponds to faster 
peptidyl transferase reaction. Therefore, the trend of variation of 
the most probable dwell time with $\omega_{p}$ is consistent with 
the intuitive expectation that the slower is the peptidyl transferase 
reaction, the longer is the dwell time.  

The kinetic parameter $\omega_{r2}$ is a measure of the rate of 
rejection of the aa-tRNA by kinetic proofreading. Consequently, 
the effect of the variation of the $\omega_{r2}$ on the most 
probable dwell time is opposite to that of $\omega_{p}$ (see  
Fig.\ref{fig-ftomegar2}); the higher is the frequency of tRNA 
rejection by kinetic proofreading, the longer is the dwell time.

\section{Mean rate of polymerization: a Michaelis-Menten-like equation} 
\label{sec-mmeqn}

The average Dwell time can be calculated by substituting (\ref{eq-ftfinal}) 
into the definition 
\begin{equation}
\langle t \rangle = \int_0^{+\infty} tf(t) \,dt.
\end{equation}
Hence, the expression for $\langle t \rangle$ 
\begin{equation}
\langle t \rangle=\dfrac{1}{\omega_1}+\dfrac{1}{\omega_2}+\dfrac{1}{\omega_3}+\dfrac{\omega_p}{\omega_p+\Omega_p}\biggl(\dfrac{1}{\omega_4}+\dfrac{1}{\omega_5}\biggr)+\dfrac{\Omega_p}{\omega_p+\Omega_p}\biggl(\dfrac{1}{\Omega_4}+\dfrac{1}{\Omega_5}\biggr)
\label{eq-avteqn}
\end{equation}
written in terms of 
$\omega_{1},\omega_{2},\omega_{3},\omega_{4},\omega_{5},\Omega_{4}$ and $\Omega_{5}$ 
has a clear physical meaning. In the special case $\Omega_{p} = 0$, 
$\langle t \rangle = \sum_{i=1}^{5} \omega_{i}^{-1}$. However, if 
$\Omega_{p} \neq 0$, then the sum $\omega_{4}^{-1}+\omega_{5}^{-1}$ 
is replaced by the last two terms of (\ref{eq-avteqn}) where 
$\omega_{p}/(\omega_{p}+\Omega_{p})$ and 
$\Omega_{p}/(\omega_{p}+\Omega_{p})$ are 
the weight factors associated with the two paths emanating from the 
state labelled by $3$.

Finally, in terms of the rate constants for the mechano-chemical 
transitions in the model depicted in fig.\ref{fig-5state} 
\begin{eqnarray}
\langle t \rangle &=& \frac{1}{V}= \frac{1}{\omega_a}\biggl(1+\frac{\omega_{r1}}{\omega_{h1}}\biggr)\biggl(1+\frac{\omega_{r2}}{\omega_p+\Omega_p}\biggr)+\frac{1}{\omega_{h1}}\biggl(1+\frac{\omega_{r2}}{\omega_p+\Omega_p}\biggr)+\frac{1}{\omega_p+\Omega_p} \nonumber \\
&+&\frac{\omega_p}{\omega_p+\Omega_p}\biggl[\frac{1}{\omega_{bf}}\biggl(1+\frac{\omega_{br}}{\omega_{h2}}\biggr)+\frac{1}{\omega_{h2}}\biggr]+\frac{\Omega_p}{\omega_p+\Omega_p}\biggl[\frac{1}{\Omega_{bf}}\biggl(1+\frac{\Omega_{br}}{\Omega_{h2}}\biggr)+\frac{1}{\Omega_{h2}}\biggr]\nonumber\\
\label{eq-avtfinal}
\end{eqnarray}

Next we'll show that the equation (\ref{eq-avtfinal}) can be 
re-expressed in a form that resembles the Michaelis-Menten 
equation for simple enzymatic reactions \cite{dixon}. Consider  
an enzymatic reaction of the type 
\begin{equation}
E + S \mathop{\rightleftharpoons}^{\omega_{+1}}_{\omega_{-1}} ES \mathop{\rightarrow}^{\omega_{+2}} E + P
\label{eq-MMscheme} 
\end{equation}
where $E$ is the enzyme, $S$ is the substrate and $P$ is the 
product of the reaction catalyzed by $E$. Given that the total 
initial concentration of the enzyme is $[E]_{0}$, the rate 
$V$ of this reaction is given by 
\begin{equation}
\frac{1}{V} =\frac{1}{V_{max}}+\frac{K_M}{V_{max}} \frac{1}{[S]}
\end{equation}
where the maximum possible rate of the reaction is 
\begin{equation}
V_{max} = \omega_{+2} [E]_{0} \equiv \tilde{\omega}_{2}
\end{equation}
and the Michaelis constant $K_M$ is given by 
\begin{equation}
K_M=\frac{\omega_{-1}+\omega_{+2}}{\omega_{+1}}
\end{equation}

In the case of our model, we assume that the ``pseudo'' first 
order rate constant $\omega_{a}$ can be written as 
$\omega_a = \omega_a^0 [tRNA]$ where $[tRNA]$ is the concentration 
of tRNA molecules in the solution. Treating tRNA molecules as 
the analogues of the substrates in an enzymatic reaction, 
equation (\ref{eq-avtfinal}) can be re-expressed as a 
Michaelis-Menten-like Equation \cite{dixon}
\begin{equation}
\langle t \rangle =\frac{1}{V_{max}}+\frac{K_M}{V_{max}} \frac{1}{[tRNA]}
\end{equation}
where
\begin{eqnarray}
\frac{1}{V_{max}} &=& \frac{1}{\omega_2^{eff}}= \frac{1}{\omega_{h1}}\biggl(1+\frac{\omega_{r2}}{\omega_p+\Omega_p}\biggr)+\frac{1}{\omega_p+\Omega_p}\nonumber\\
&+&\frac{\omega_p}{\omega_p+\Omega_p}\biggl[\frac{1}{\omega_{bf}}\biggl(1+\frac{\omega_{br}}{\omega_{h2}}\biggr)+\frac{1}{\omega_{h2}}\biggr] +\frac{\Omega_p}{\omega_p+\Omega_p}\biggl[\frac{1}{\Omega_{bf}}\biggl(1+\frac{\Omega_{br}}{\Omega_{h2}}\biggr)+\frac{1}{\Omega_{h2}}\biggr]\nonumber\\
\label{a2}
\end{eqnarray}
and the Michaelis constant  
\begin{equation}
K_M=\frac{\omega_{-1}^{eff}+\omega_{2}^{eff}}{\omega_a^0}
\end{equation}
with
\begin{equation}
\omega_{-1}^{eff}=\biggl(\frac{\omega_{r1}}{\omega_{h1}}+\frac{\omega_{r2}}{\omega_p+\Omega_p}+\frac{\omega_{r1}\omega_{r2}}{\omega_{h1}(\omega_p+\Omega_p)}\biggr)\omega_2^{eff}
\end{equation}
Thus, the mean dwell time for the ribosomes follows a 
Michaelis-Mentan-like equation. This result is consistent 
with the experimental observations in recent years 
\cite{qian02,english06,kou05,min05,min06,basu09}
that, in spite of the fluctuations of an enzymatic reaction 
catalyzed by a {\it single} enzyme molecule, the average rate 
of the reaction is, most often, given by the Michaelis-Menten 
equation. What makes the Michaelis-Menten-like equation for 
the ribosome even more interesting is the fact that a single 
ribosome is not a single enzyme molecule, but it provides a 
platform for the coordinated operation of several bio-catalysts.

\section{Effects of crowding on the dwell time distribution} 
\label{sec-crowding}

It is well known  that most often a large number of ribosomes 
simultaneously move on the same mRNA track each polymerizing 
one copy of the same protein. This phenomenon is usually referred 
to as ribosome traffic because of its superficial similarity with 
vehicular traffic on highways 
\cite{macdonald68,macdonald69,lakatos03,shaw03,shaw04a,shaw04b,chou03,chou04,dong1,dong2,cook,ciandrini,basu07,mitarai08}
. Suppose, ${\ell}$ denotes the number of codons that a ribosome 
can cover simultaneously. Extending the prescription used in our 
earlier works on ribosome traffic \cite{basu07} for capturing the 
steric interactions of the ribosomes, we replace the 
equations (\ref{eq-m5}), (\ref{eq-m7} and (\ref{eq-m8}) by 
\begin{equation}
\frac{dP_5(i,t)}{dt}=\omega_{bf} P_4(i,t)- \omega_{br} P_5(i,t) - \omega_{h2} P_5(i,t) Q(\underbar{i}|i+{\ell})
\label{eq-m5traf}
\end{equation}
\begin{equation}
\frac{dP_5^{*}(i,t)}{dt}=\Omega_{bf} P_4^{*}(i,t)- \Omega_{br} P_5^{*}(i,t) - \Omega_{h2} P_5^{*}(i,t) Q(\underbar{i}|i+{\ell})
\label{eq-m7traf}
\end{equation}
\begin{equation}
\frac{d\widetilde{P_1}(i,t)}{dt}=\biggl(\omega_{h2}P_5(i-1,t) + \Omega_{h2}P_5^*(i-1,t)\biggr) Q(\underbar{i-1}|i-1+\ell)
\label{eq-m8traf}
\end{equation}
where $Q(\underbar{i}|j) = 1 - P(\underbar{i}|j)$ is the conditional
probability that, given a ribosome in site i, site j is empty. All 
the other equations for $P_{\mu}(i,t)$ remain unchanged. Note that 
these equations have been written under the {\it mean-field approximation}. 

In the limit $L \to \infty$, all the sites are treated on the same 
footing so that the site-dependence of $P_{\mu}(i,t)$ drops out. 
In this limit, $Q$ takes the simple form \cite{basu07}
\begin{eqnarray} 
Q(\underbar{i}|i+{\ell})= \frac{1-\rho{\ell}}{1+\rho-\rho{\ell}}.
\label{eq-qpbc}
\end{eqnarray}
where $\rho$ is the number density of the ribosomes (i.e., number of 
ribosomes per unit length of the mRNA track). 

Therefore, in ribosome traffic the distribution of the dwell times 
of the ribosomes is given by the expression (\ref{eq-ftfinal}) where 
$\omega_{h2}$ and $\Omega_{h2}$ are replaced by $\omega_{h2} Q$ and 
$\Omega_{h2}Q$, respectively. Using the expression (\ref{eq-qpbc}), 
we get the DTD for the given number density $\rho$ of the ribosomes. 
For the purpose of graphical demostration of the effects of ribosome 
crowding on the DTD, we use the {\it coverage} density 
$\rho_{cov} = \rho ~\ell$. In Fig.\ref{fig-ftrho} we plot the DTD for 
a few different values of $\rho_{cov}$. The higher is the density, the 
stronger is the hindrance and longer is the dwell time. Moreover, 
higher coverage density introduces stronger correlations which our 
mean-field equations ignore. Consequently, our analytical predictions, 
based on equations written under the mean-field approximation, deviate 
more and more from the corresponding simulation data as the coverage 
density increases.

\section{Comparison with experimental data} 
\label{sec-experiment}

The distribution of dwell times involves essentially five different 
rate-determining parameters $\omega_i$ ($i=1,2,3,4,5$) if the 
translational fidelity is perfect. This is consistent with the 
earlier observation \cite{wen09} that the best fit to the simulation 
data was obtained with five rate-determining parameters. However, 
the dwell time distribution obtained from the single-molecule 
experiments (more precisely, single-ribosome manipulation) fit 
reasonably well with a difference of just two exponentials 
\cite{wen08}. This strongly indicates the possibility that only 
two of the five rate-determining parameters were rate-limiting under 
the conditions maintained in those experiments. However, for a 
quantitative testing of the predictions of our theoretical model,  
one should use a mRNA template with a homogeneous codon sequence and 
only two species of aa-tRNA one of which is cognate (corresponding 
to the codons in the coding sequence of the mRNA) and the other 
is either near-cognate or non-cognate. 

\begin{figure}[t]
\begin{center}
\includegraphics[angle=-90,width=0.6\columnwidth]{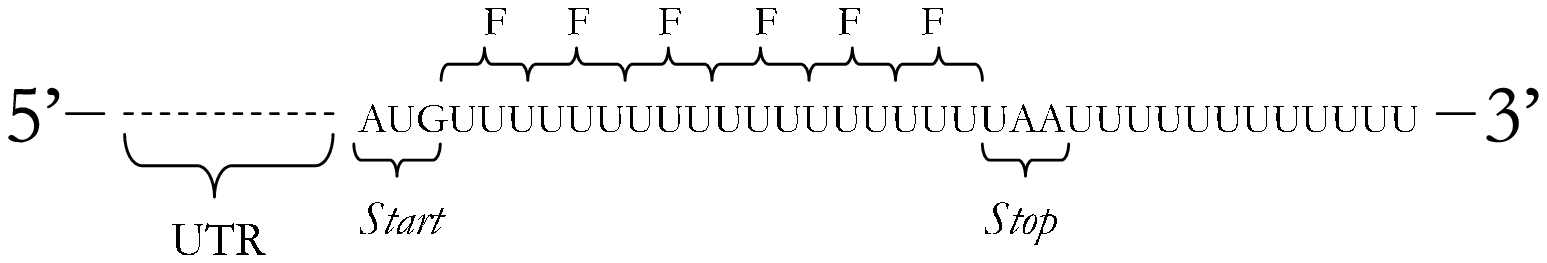}
\end{center}
\caption{A schematic description of a mRNA with homogeneous (poly-U) 
coding sequence (adapted from refs.\cite{uemura10,aitken10}).  
}
\label{fig-expt}
\end{figure}

In order to clarify the proposed experimental set up, let us consider 
the concrete example shown schematically in fig. \ref{fig-expt}. 
This example is essentially the protocols used by Uemura et al.  
\cite{uemura10} in their single-molecule studies of translation in 
real time. The actual coding sequence to be translated consists of 
$n_c$ number of identical codons; in fig.\ref{fig-expt} $n_c=6$ and 
each codon is UUU which codes for the amino acid Phenylalanine 
(abbreviated {\it Phe} or {\bf F}). The coding sequence is preceeded 
by a start codon AUG and is followed by a stop codon UAA. The start 
codon itself is preceeded by an untranslated region (UTR) at the 5'-end 
of the mRNA; this is required for assembling the ribosome and for 
stabilizing the pre-initiation complex. At the 3'-end, the stop 
codon is followed by a sequence of $n_{nc}$ non-coding codons 
UUU ($n_{nc}=4$ in fig.\ref{fig-expt}); this region merely 
ensures the absence of any ``edge effect'', i.e., the translation 
is not affected when the ribosome approaches the 3'-end of the 
codong sequence. A good choice for the corresponding near-cognate 
tRNA would be tRNA$^{Leu}$ because it is cognate for the codon 
CUU which codes for {\it Leucine} (abbreviated {\bf L}). 
The two distinct species of aa-tRNA should be labelled by fluorescent 
dye molecules of two different colors which could be, for example, 
Cy5 (red) and Cy3 (green). Equal concentrations of Cy5-labelled 
ternary complexe aa-tRNA$^{Phe}$-EF-Tu-GTP and Cy3-labelled 
ternary complex aa-tRNA$^{Leu}$-EF-Tu-GTP should be made available 
in the medium surrounding the ribosome to avoid any bias arising from 
difference in their concentrations. The color of the fluorescence pulse 
identifies the amino acid monomer species that elongates the polypeptide 
by one unit; monitoring the colors of the fluorescence pulses 
one would get an estimate of the translational fidelity $\phi$. 
Moreover, the time interval between the arrival of the successive  
aa-tRNA molecules provides an estimate of the dwell times of the 
ribosome. Thus, using this optical technique one would get not only 
the DTD, but also the fidelity of translation. 
Usually the coding sequence of such poly-U mRNA strands is quite 
short. Therefore, for collecting enough data to extract the DTD, the 
experiment has to be repeated sufficiently large number of times.

The effects of kinetic proofreading on the rate of translational 
error has been investigated experimentally by several different 
methods in the last three decades. However, most of those methods 
(see, for example, ref.\cite{hopfield1,hopfield2}) require bulk 
samples. But, the more recent single-molecule FRET technique used 
in the study of tRNA selection \cite{blanchard04} seems to be more 
suitable for testing our predictions on the effects of kinetic 
proofreading on the DTD.

The cluster of ribosomes translating the same mRNA simultaneously 
is usually referred to as a {\it polysome}. For studying the 
effects of ribosome crowding on the DTD, one has to measure 
simultaneously the DTD and the polysome size \cite{arava03}.

\section{Summary and conclusion} 

In this paper we have presented a theoretical model of translation that 
captures all the main steps in the mechano-chemical cycle of a ribosome 
during the elongation stage. This model also accounts for translational 
fidelity, kinetic proofreading and the crowding of the ribosomes on the 
same mRNA track. In principle, this model can be extended, by increasing 
the number of ``chemical'' states, to account for some of the sub-steps 
which have not been treated explicitly in the version of the model 
presented here. 

In spite of the details already incorporated in this model, we have 
suceeded in carrying out an {\it analytical} calculation of the 
distribution of the dwell times of the ribosomes at the successive 
codons. We have compared this theoretical estimate of the DTD  
with the corresponding numerical data which we have obtained from 
direct computer simulation of the model. If the motion of the ribosome 
is not hindered by the presence of any other ribosome on the same mRNA 
track, our analytical treatment yields the {\it exact} expression for 
the distribution for the times of its dwell at the successive codons. 
In this case, excellent agreement between theoretical prediction and 
simulation data is observed provided the data are averaged over 
sufficiently large number of samples. 

However, because of the mean-field approximation made in capturing 
the effects of crowding of the ribosomes the corresponding analytical 
expression is approximate. Therefore, the higher is the coverage 
density, the larger is the deviation of the analytical estimate from 
the simulation data which are averaged over many samples. 

We have analyzed the dependence of the DTD on some of the crucially 
important kinetic parameters of the model to elucidate the physical 
implications of the result. Our results are in good qualitative 
agreement with the experimental data reported in the literature 
\cite{wen08,wen09}. However, for the reasons explained in the 
sections \ref{sec-introduction} and \ref{sec-model}, it is not possible 
to compare these experimental data quantitatively with the analytical 
expressions of DTD which we have reported in this paper. 
We hope our theoretical predictions will stimulate further 
experimental studies along the lines suggested briefly in section 
\ref{sec-experiment} although some technical hurdles may hinder 
quick progress. 
A combination of the single-ribosome experiments and bulk measurements 
may be required for comparing the theoretically predicted variation of 
the DTD with the concentrations of tRNA, GTP and other key molecules 
involved in translation as well as with the increase of futile cycles 
and crowding.\\

It would also be desirable to extend our model in future to account  
for some important features of translation; some of these possible 
extentions are listed below. (i) Capturing the sequence heterogeneity 
of real mRNA templates will open up larger number of branched pathways 
in Fig.\ref{fig-5state}, each corresponding to a distinct species 
(cognate, non-cognate or near-cognate) of amino acid. Moreover, 
at each step of the ribosome, the rate constants for the correct and 
incorrect pathways will also depend on the codon under consideration. 
(ii) The interaction of the growing polypeptide with the exit tunnel 
in the ribosome and the spontaneous folding of the nascent protein as 
it comes out of the tunnel may affect the rate of translation. It may 
be possible to capture these effects by making some of the relevant 
rate constants, e.g., $\omega_{h2}$, $\Omega_{h2}$, etc. on the 
instantaneous length of the growing polypeptide which, in turn, is 
identical to the length of the codon sequence already translated by 
the ribosome. 
(iii) Maintaining correct reading frame on a sequence-homogeneous mRNA 
(e.g., poly-U) may be more difficult than on an inhomogeneous sequence. 
It may be desirable to extend our model allowing the possibility of 
frameshift errors \cite{farabaugh00} although, at present, the 
prescription for this extension is not obvious. 
The DTD for such a ``realistic model'' may be obtained numerically 
because an {\it analytical} derivation of the general expression may 
not be possible in the forseeable future. \\

\noindent {\bf Acknowledgements:} One of the authors (DC) thanks 
Joachim Frank and Joseph D. Puglisi for useful correspondences.

\noindent {\bf Appendix}

Solving the equations (\ref{eq-m1})-(\ref{eq-m7}) under the initial 
condition (\ref{eq-initial}) we get  
\begin{equation}
P_1(t)=C_1 e^{-\omega_1 t}+C_2 e^{-\omega_2 t}+ C_3 e^{-\omega_3 t}
\label{eq-P1}
\end{equation}

\begin{equation}
P_2(t)=\dfrac{C_1 \omega_a e^{-\omega_1 t}}{\omega_{r1}+\omega_{h1}-\omega_1}+\dfrac{C_2 \omega_a e^{-\omega_2 t}}{\omega_{r1}+\omega_{h1}-\omega_2}+\dfrac{C_3 \omega_a e^{-\omega_3 t}}{\omega_{r1}+\omega_{h1}-\omega_3}
\label{eq-P2}
\end{equation}
\begin{eqnarray}
P_3(t) &=& \dfrac{C_1 \omega_{h1}\omega_a e^{-\omega_1 t}}{(\omega_{r1}+\omega_{h1}-\omega_1)(\omega_{r2}+\omega_p+\Omega_p-\omega_1)}+\dfrac{C_2 \omega_{h1} \omega_{a} e^{-\omega_2 t}}{(\omega_{r1}+\omega_{h1}-\omega_2)(\omega_{r2}+\omega_p+\Omega_p-\omega_2)} \nonumber \\
&+&\dfrac{C_3 \omega_{h1}\omega_a e^{-\omega_3 t}}{(\omega_{r1}+\omega_{h1}-\omega_3)(\omega_{r2}+\omega_p+\Omega_p-\omega_3)}
\label{eq-P3}
\end{eqnarray}
\begin{eqnarray}
P_4(t)&=& \dfrac{C_1 \omega_p\omega_{h1}\omega_a (\omega_{h2}+\omega_{br}-\omega_1) e^{-\omega_1 t}}{(\omega_{r1}+\omega_{h1}-\omega_1)(\omega_{r2}+\omega_p+\Omega_p-\omega_1)(\omega_4-\omega_1)(\omega_5-\omega_1)}\nonumber \\
&+&\dfrac{C_2\omega_p \omega_{h1} \omega_{a}(\omega_{h2}+\omega_{br}-\omega_2)e^{-\omega_2 t}}{(\omega_{r1}+\omega_{h1}-\omega_2)(\omega_{r2}+\omega_p+\Omega_p-\omega_2)(\omega_4-\omega_2)(\omega_5-\omega_2)} \nonumber \\
&+&\dfrac{C_3 \omega_p\omega_{h1}\omega_a(\omega_{h2}+\omega_{br}-\omega_3)e^{-\omega_3 t}}{(\omega_{r1}+\omega_{h1}-\omega_3)(\omega_{r2}+\omega_p+\Omega_p-\omega_3)(\omega_4-\omega_3)(\omega_5-\omega_3)}\nonumber \\
&+&C_4 e^{-\omega_4 t}+C_5 e^{-\omega_5 t}\nonumber\\
\label{eq-P4}
\end{eqnarray}
\begin{eqnarray}
P_5(t)&=& \dfrac{C_1 \omega_{bf}\omega_p\omega_{h1}\omega_a e^{-\omega_1 t}}{(\omega_{r1}+\omega_{h1}-\omega_1)(\omega_{r2}+\omega_p+\Omega_p-\omega_1)(\omega_4-\omega_1)(\omega_5-\omega_1)}\nonumber \\
&+&\dfrac{C_2\omega_{bf}\omega_p \omega_{h1} \omega_{a} e^{-\omega_2 t}}{(\omega_{r1}+\omega_{h1}-\omega_2)(\omega_{r2}+\omega_p+\Omega_p-\omega_2)(\omega_4-\omega_2)(\omega_5-\omega_2)} \nonumber \\
&+&\dfrac{C_3\omega_{bf} \omega_p\omega_{h1}\omega_a e^{-\omega_3 t}}{(\omega_{r1}+\omega_{h1}-\omega_3)(\omega_{r2}+\omega_p+\Omega_p-\omega_3)(\omega_4-\omega_3)(\omega_5-\omega_3)}\nonumber \\
&+&\dfrac{C_4 \omega_{bf} e^{-\omega_4 t}}{\omega_{h2}+\omega_{br}-\omega_4}+\dfrac{C_5\omega_{bf} e^{-\omega_5 t}}{\omega_{h2}+\omega_{br}-\omega_5}
\label{eq-P5}
\end{eqnarray}
\begin{eqnarray}
P_4^*(t)&=& \dfrac{C_1 \Omega_p\omega_{h1}\omega_a (\Omega_{h2}+\Omega_{br}-\omega_1)e^{-\omega_1 t}}{(\omega_{r1}+\omega_{h1}-\omega_1)(\omega_{r2}+\omega_p+\Omega_p-\omega_1)(\Omega_4-\omega_1)(\Omega_5-\omega_1)}\nonumber \\
&+&\dfrac{C_2\Omega_p \omega_{h1} \omega_{a}(\Omega_{h2}+\Omega_{br}-\omega_2)e^{-\omega_2 t}}{(\omega_{r1}+\omega_{h1}-\omega_2)(\omega_{r2}+\omega_p+\Omega_p-\omega_2)(\Omega_4-\omega_2)(\Omega_5-\omega_2)} \nonumber \\
&+&\dfrac{C_3 \Omega_p\omega_{h1}\omega_a(\Omega_{h2}+\Omega_{br}-\omega_3)e^{-\omega_3 t}}{(\omega_{r1}+\omega_{h1}-\omega_3)(\omega_{r2}+\omega_p+\Omega_p-\omega_3)(\Omega_4-\omega_3)(\Omega_5-\omega_3)}\nonumber \\
&+&D_4 e^{-\Omega_4 t}+D_{5} e^{-\Omega_5 t}\nonumber\\
\label{eq-P4star}
\end{eqnarray}
\begin{eqnarray}
P_5^*(t)&=& \dfrac{C_1 \Omega_{bf}\Omega_p\omega_{h1}\omega_a e^{-\omega_1 t}}{(\omega_{r1}+\omega_{h1}-\omega_1)(\omega_{r2}+\omega_p+\Omega_p-\omega_1)(\Omega_4-\omega_1)(\Omega_5-\omega_1)}\nonumber \\
&+&\dfrac{C_2\Omega_{bf}\Omega_p \omega_{h1} \omega_{a} e^{-\omega_2 t}}{(\omega_{r1}+\omega_{h1}-\omega_2)(\omega_{r2}+\omega_p+\Omega_p-\omega_2)(\Omega_4-\omega_2)(\Omega_5-\omega_2)} \nonumber \\
&+&\dfrac{C_3\Omega_{bf} \Omega_p\omega_{h1}\omega_a e^{-\omega_3 t}}{(\omega_{r1}+\omega_{h1}-\omega_3)(\omega_{r2}+\omega_p+\Omega_p-\omega_3)(\Omega_4-\omega_3)(\Omega_5-\omega_3)}\nonumber \\
&+&\dfrac{D_4 \Omega_{bf} e^{-\Omega_4 t}}{\Omega_{h2}+\Omega_{br}-\Omega_4}+\dfrac{D_{5}\Omega_{bf} e^{-\Omega_5 t}}{\Omega_{h2}+\Omega_{br}-\Omega_5}
\label{eq-P5star}
\end{eqnarray}
where $\omega_{1}$, $\omega_{2}$ and $\omega_{3}$ are the solutions 
of the cubic equation (\ref{eq-w123}) while $\omega_{4}$, $\omega_{5}$ 
are the solutions of the quadratic equation (\ref{eq-w45}) and 
$\Omega_{4}$, $\Omega_{5}$ are the solutions of the quadratic equation 
(\ref{eq-w45star}).   
The coefficients $C_{\nu}$ ($\nu=1,2,...,5$) and $D_{4}$, $D_{5}$, 
which are determined by the initial condition (\ref{eq-initial}), 
are as follows: 
\begin{equation}
C_1=\dfrac{(\omega_{r1}+\omega_{h1}-\omega_1)(\omega_{r2}+\omega_p+\Omega_p-\omega_1)}{(\omega_2-\omega_1)(\omega_3-\omega_1)}
\end{equation}
\begin{equation}
C_2=\dfrac{(\omega_{r1}+\omega_{h1}-\omega_2)(\omega_{r2}+\omega_p+\Omega_p-\omega_2)}{(\omega_3-\omega_2)(\omega_1-\omega_2)}
\end{equation}
\begin{equation}
C_3=\dfrac{(\omega_{r1}+\omega_{h1}-\omega_3)(\omega_{r2}+\omega_p+\Omega_p-\omega_3)}{(\omega_1-\omega_3)(\omega_2-\omega_3)}
\end{equation}
\begin{equation}
C_4=\dfrac{(\omega_{h2}+\omega_{br}-\omega_4)\omega_p\omega_{h1}\omega_a}{(\omega_5-\omega_4)(\omega_1-\omega_4)(\omega_2-\omega_4)(\omega_3-\omega_4)}
\end{equation}
\begin{equation}
C_5=\dfrac{(\omega_{h2}+\omega_{br}-\omega_5)\omega_p\omega_{h1}\omega_a}
{(\omega_1-\omega_5)(\omega_2-\omega_5)(\omega_3-\omega_5)(\omega_4-\omega_5)}
\end{equation}
\begin{equation}
D_4=\dfrac{(\Omega_{h2}+\Omega_{br}-\Omega_4)\Omega_p\omega_{h1}\omega_a}{(\Omega_5-\Omega_4)(\omega_1-\Omega_4)(\omega_2-\Omega_4)(\omega_3-\Omega_4)} 
\end{equation}
\begin{equation}
D_{5}=\dfrac{(\Omega_{h2}+\Omega_{br}-\Omega_5)\Omega_p\omega_{h1}\omega_a}{(\omega_1-\Omega_5)(\omega_2-\Omega_5)(\omega_3-\Omega_5)(\Omega_4-\Omega_5)}
\end{equation}


\end{document}